\documentclass[aps,prl,showpacs,superscriptaddress,groupedaddress]{revtex4}

\usepackage[dvipdfmx]{graphicx}
\usepackage{dcolumn}
\usepackage{bm}
\usepackage{amsmath}
\usepackage{amsthm} 
\usepackage{threeparttable}
\usepackage{color}
\usepackage{mathrsfs}
\def\v#1{{\boldsymbol{#1}}} 
\def\m#1{{\mathcal{#1}}} 

\begin{document}


\title{Enhanced axial migration of a deformable capsule in pulsatile channel flows}


\author{Naoki Takeishi}
\email{ntakeishi@kit.ac.jp}
\affiliation{Department of Mechanical Engineering, Kyoto Institute of Technology, Goshokaido-cho, Matsugasaki, Sakyo-ku, Kyoto, 606-8585, Japan}
\affiliation{Graduate School of Engineering Science, Osaka University, 1-3 Machikaneyama, Toyonaka, Osaka, 560-8531, Japan.}

\author{Marco Edoardo Rosti}
\email{marco.rosti@oist.jp}
\affiliation{Complex Fluids and Flows Unit, Okinawa Institute of Science and Technology Graduate University, 1919-1 Tancha, Onna-son, Okinawa 904-0495, Japan.}


\date{First submission \today}

\begin{abstract}
We present numerical analysis of the lateral movement of a deformable spherical capsule in a pulsatile channel flow, with a Newtonian fluid in almost inertialess condition and at a small confinement ratio $a_0/R$ = 0.4, where $R$ and $a$ are the channel and capsule radius. We find that the speed of the axial migration of the capsule can be accelerated by the flow pulsation at a specific frequency. The migration speed increases with the oscillatory amplitude, while the most effective frequency remains basically unchanged and independent of the amplitude.
Our numerical results form a fundamental basis for further studies on cellular flow mechanics, since pulsatile flows are physiologically relevant in human circulation, potentially affecting the dynamics of deformable particles and red blood cells (RBCs), and can also be potentially exploited in cell focusing techniques.
\end{abstract}


\maketitle


\section{I. Introduction}
High-throughput measurements of single-cell behaviour under confined channel flow is of fundamental importance and technical requirement in bioengineering applications such as cellular-level diagnoses for blood diseases. Although several attempts have addressed this issue and gained insights into (soft) particle dynamics in microchannels~\citep{Ciftlik2013, Fregin2019, Ito2017}, cell manipulation including label-free cell alignment, sorting, and separation still face major challenges. Along with the aforementioned experimental studies, recent numerical simulations revealed the mechanical background regarding the lateral movement of particles, e.g. in~\citep{Alghalibi2019, Takeishi2021, Takeishi2022JFM}. The lateral movement of deformable spherical particles in almost inertialess conditions was originally reported in~\citet{Karnis1963}, and these results have been the fundamental basis to describe the phenomena observed in microfluidics~\citep{Kim2019} but also in {\it in vivo} microcirculations~\citep{Secomb2017}. In particular, it was found that a deformable spherical particle tends to move towards the channel axis and settles there. Hereafter we will call this phenomenon as ``axial migration".

It is known that the presence of axial migration or non-axial migration depends on particle shape and initial orientation angles. An RBC modelled as a biconcave capsule does not always exhibit axial migration especially in the tank-treading slipper shape, obtained with high $Ca$ and high $\lambda$~\citep{Guckenberger2018, Takeishi2021}. 
Furthermore, RBCs have bistable flow mode, so-called rolling and tumbling motions, which depend on the initial cell orientations~\citep{Takeishi2022JFM}. Thus, the original spherical shape is one of requirements for the axial migration in (almost) inertialess conditions.
In a recent work, the framework of the axial migration of a droplet has been extended by~\citet{Santra2021} by including the effect of an electric field, and finding that as the strength of the electric field increases, droplets can reach the centreline at a faster rate with reduced axial oscillations. Furthermore, a deformation-dependent propulsion of soft particles, including biological cells, were confirmed experimentally by~\citet{Krauss2022} and numerically by~\citet{Schmidt2022}.

Despite these efforts, the effect of a pulsatile flow on the axial migration of capsules 
has not been described and understood yet. The objective of this study is thus to clarify whether frequency-dependent axial migration of the spherical capsule occurs in confined channel flows. More precisely, \textit{can the time necessary for the axial migration be controlled by the channel pulsations? Is there an optimal pulsation frequency to do that?} As we will describe in the following, our investigation of the capsule dynamic show that \textit{it indeed exists an optimal frequency to speed-up the capsule axial migration by up to $80\%$ in the range of parameters investigated here.}
\begin{figure}[htbp]
  \centering
    \includegraphics[clip,width=10cm]{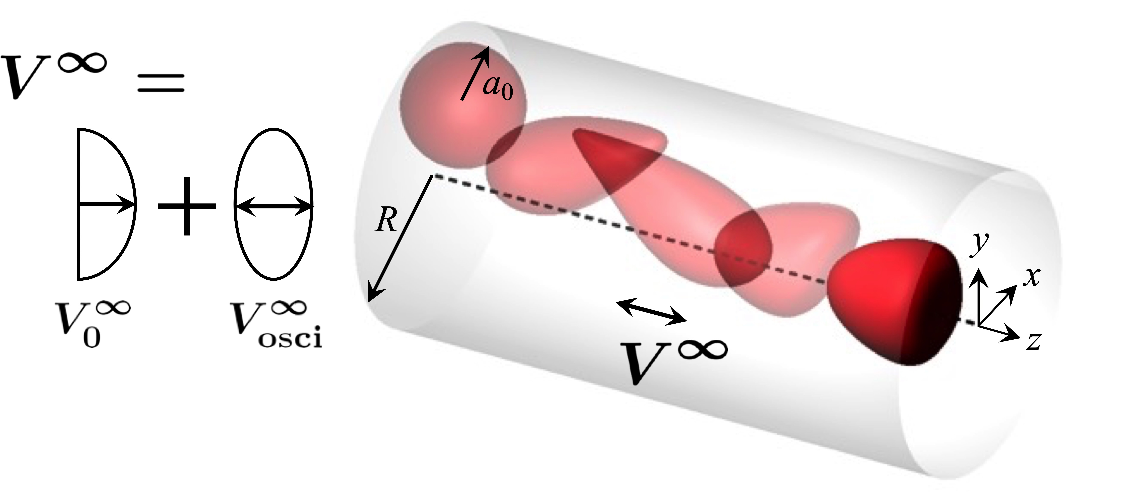}
    \caption{
    Visualization of a spherical capsule with radius $a_0$ in a tube with radius of $R$ under a pulsatile flow with velocity $V^\infty$, 
    which can be decomposed into the steady parabolic flow $V_0^\infty$ and the oscillatory flow $V_\mathrm{osci}^\infty$ in the absence of any cells.
    The capsule, initially placed near the wall, exhibits axial migration.
    }
    \label{fig:snap_shot}
\end{figure}

\section{II. Problem statement and methods}
\subsection{A. Problem statement and governing equations}
To answer these fundamental questions, we perform a series of fully resolved numerical simulations. We consider the motion of an initially spherical capsule with diameter $d_0$ (= $2a_0$ = $8$ $\mu$m) flowing in a circular channel of diameter $D$ (= 2$R$ = $20$ $\mu$m), see Fig.~\ref{fig:snap_shot}.
The capsule is made by an elastic membrane, separating two Newtonian fluids, which satisfy the incompressible Navier--Stokes equations, and have the same density $\rho$ but different viscosity (inside) $\mu_1$ and (outside) $\mu_0$.
The membrane is modeled as an isotropic and hyperelastic material following the Skalak constitutive (SK) law~\citep{Skalak1973}. In particular, the strain energy $w$ of the SK law is given by
\begin{equation}
  w = \frac{G_s}{4} \left( I_1^2 + 2I_1 - 2I_2 + C I_2^2\right),
  \label{SK}
\end{equation}
where $G_s$ is the surface shear elastic modulus, $C$ is a dimensionless material coefficient that measures the resistance to the area dilation, $I_1 (= \lambda_1^2 + \lambda_2^2 - 2)$ and $I_2 (= \lambda_1^2 \lambda_2^2 -1 = J_s^2 - 1)$ are the first and second invariants of the Green-Lagrange strain tensor, $\lambda_i$ ($i = 1$ and $2$) are the two principal in-plane stretch ratios, and $J_s = \lambda_1 \lambda_2$ is the Jacobian, which expresses the ratio of the deformed to reference surface areas. The area dilation modulus of the SK law is $K_s = G_s \left( 1 + 2C \right)$~\cite{BarthesBiesel2002}. Bending resistance is also considered~\citep{Li2005}, with a bending modulus $k_b = 5.0 \times 10^{-19}$ J~\citep{Puig-de-Morales-Marinkovic2007}.
In this study, the surface shear elastic modulus is determined to be $G_s$ = 4 $\mu$N/m to mimic the value found in human RBCs~\cite{Takeishi2014, Takeishi2019JFM}. Assuming the area incompressibility of the membrane and also following previous study by~\citep{BarthesBiesel2002}, we set as $C = 10^2$. These membrane parameters successfully captured the characteristic stable deformation and dynamics of RBCs both in single and multi-cellular interaction problems~\cite{Takeishi2014, Takeishi2019JFM}.

Neglecting inertial effects on the membrane deformation, the static local equilibrium equation of the membrane is given by
\begin{equation}
  \nabla_s \cdot \v{T} + \v{q} = \v{0},
  \label{StrongForm}
\end{equation}
where $\nabla_s (= \left( \v{I} - \v{n} \v{n} \right) \cdot \nabla)$ is the surface gradient operator, $\v{n}$ is the unit normal outward vector in the deformed state, $\v{q}$ is the load on the membrane, and $\v{T}$ is the in-plane elastic tension that is obtained from the SK law~\eqref{SK}.

The two fluids separated by the membrane are governed by the incompressible Navier--Stokes equations,
\begin{align}
	\rho \left( \frac{\partial \v{v}}{\partial t} + \v{v} \cdot \nabla \v{v} \right)
	&= \nabla \cdot \v{\sigma}^f  + \rho \v{f}, \\
	\nabla \cdot \v{v} &= 0,
\end{align}
where
\begin{align}
	\v{\sigma}^f = -p\v{I} + \mu \left( \nabla \v{v} + \nabla \v{v}^T \right).
\end{align}
In the previous equations, $\v{\sigma}^f$ is the total stress tensor of the flow, $p$ is the pressure, $\rho$ is the fluid density,$\v{f}$ is the body force, and $\mu$ is the viscosity of the liquids, which is expressed using the volume fraction of the inner fluid $\alpha$ (0 $\leq \alpha \leq$ 1) as:
\begin{align}
        \mu = \left\{ 1 + \left( \lambda - 1 \right) \alpha \right\} \mu_0.
\end{align}
The dynamic condition requires that the load $\v{q}$ is equal to the traction jump $\left( \v{\sigma}^f_{out} - \v{\sigma}^f_{in} \right)$ across the membrane:
\begin{align}
	\v{q} = \left( \v{\sigma}^f_\mathrm{out} - \v{\sigma}^f_\mathrm{in} \right) \cdot \v{n},
\end{align}
where the subscripts `out' and `in' represent the outer and internal regions of the capsule.

The flow in the channel is sustained by a uniform pressure gradient $\nabla p_0$, which can be related to the maximum fluid velocity in the channel as $\nabla p_0 = - 4 \mu_0 V_\mathrm{max}^\infty/R^2$. The pulsation is instead given by a superimposed sinusoidal function, such that the total pressure gradient is
\begin{equation}
  \nabla p (t) = \nabla p_0 + \left( \nabla p^\mathrm{amp} \right) \sin{(2 \pi f t)}.
\end{equation}
The problem is governed by six main non-dimensional numbers: \textit{i}) the Reynolds number $Re = \rho D V_\mathrm{max}^\infty/\mu_0$; 
\textit{ii}) the capillary number $Ca = \mu_0 \dot{\gamma}_\mathrm{m} a_0/G_s$, where $\dot{\gamma}_\mathrm{m} = V_\mathrm{max}^{\infty}/4 R$; \textit{iii}) the viscosity ratio between the two fluids $\lambda = \mu_1/\mu_0$; \textit{iv}) the confinement ratio $a_0/R$; \textit{v}) the non-dimensional pulsation frequency $f^\ast = f/\dot{\gamma}_\mathrm{m}$; \textit{vi}) the non-dimensional pulsation amplitude $\nabla p^\mathrm{amp} / \nabla p_0$ . In this work, all simulations are performed in an almost inertialess condition, keeping the Reynolds number low and fixed to the value $Re = 0.2$; also, we limit our main analysis to a confinement ratio of $0.4$. In the Appendix~\S A
we verify the sensitivity of the results to these two parameters [See Fig.~\ref{fig:verification}].
Instead here we comprehensively vary the amplitude and frequency of the pulsation, the viscosity ratio and the capillary number.

\subsection{B. Numerical methods}
The governing equations for the fluid are discretised by the lattice Boltzmann method (LBM) based on the D3Q19 model~\citep{Chen1998}. We track the Lagrangian points of the membrane material points $\v{x}(\v{X},t)$ over time, where $\v{X}$ is a material point on the membrane in the reference state. Based on the virtual work principle, the above strong-form equation (\ref{StrongForm}) can be rewritten in weak form as 
\begin{equation}
  \int_S \v{\hat{u}} \cdot \v{q} dS = \int_S \v{\hat{\epsilon}} : \v{T} dS,
  \label{WeakForm}
\end{equation}
where $\v{\hat{u}}$ and $\v{\hat{\epsilon}} = ( \nabla_s \v{\hat{u}} + \nabla_s \v{\hat{u}}^T )\big/2$ are the virtual displacement and virtual strain, respectively. The finite element method (FEM) is used to solve equation (\ref{WeakForm}) and obtain the load $\v{q}$ acting on the membrane~\citep{Walter2010}. The velocity at the membrane node is obtained by interpolating the velocities at the fluid node using the immersed boundary method~\citep{Peskin2002}. The membrane node is updated by Lagrangian tracking with the no-slip condition. The explicit fourth-order Runge--Kutta method is used for the time integration. 
The volume-of-fluid method~\citep{Yokoi2007} and front-tracking method~\citep{Unverdi1992} are employed to update the viscosity in the fluid lattices. A volume constraint is implemented to counteract the accumulation of small errors in the volume of the individual cells~\citep{Freund2007}: in our simulation, the volume error is always maintained lower than $1.0 \times 10^{-3}$\%, as tested and validated in our previous study of cell flow in circular channels~\citep{Takeishi2016}.
For further details of the methods we refer to our previous work~\citep{Takeishi2019JFM, Takeishi2022JFM}. 

Periodic boundary conditions are imposed in the flow direction ($z$-direction, see also Fig.~\ref{fig:snap_shot} and Fig.~\ref{fig:snap_timehist}b). No-slip conditions are employed for the walls (radial direction). The mesh size of the LBM for the fluid was set to be $250$ nm, and that of the finite elements describing the membrane was approximately $250$ nm (an unstructured mesh with $5,120$ elements was used for the FEM). Overall, we use a resolution of $32$ fluid lattices per diameter of the capsule. The chosen resolution has been shown in the past to successfully represent single- and multi-cellular dynamics~\citep{Takeishi2014, Takeishi2019JFM, Takeishi2021}.

\section{III. Results and discussion}
\begin{figure}[htbp]
  \centering
    \includegraphics[clip,width=17cm]{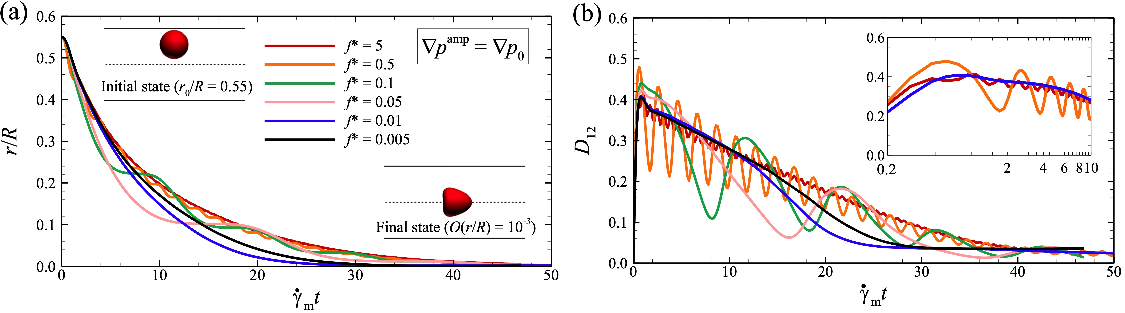}
    \caption{
    Time history of (a) the radial position of the capsule centroid $r/R$ and (b) time history of the Taylor parameter $D_{12}$ for different non-dimensional frequency $f^\ast$. The inset images in panel (a) represent the capsule initial state ($r_0/R = 0.55$ at $\dot{\gamma}_\mathrm{m} t = 0$) and the final stable state at the channel center line ($r/R \approx 0 $ at $\dot{\gamma}_\mathrm{m} t = 50$).
    All the results are obtained with $\nabla p^{amp} = \nabla p_0$, $Ca = 1.2$, and $\lambda = 1$.
    }
    \label{fig:snap_timehist}
\end{figure}

First, we investigate the trajectory of the capsule centroids for different frequencies $f^\ast = f/\dot{\gamma}_\mathrm{m}$. The time history of the radial position of the capsule centroid $r$ is shown in Fig.~\ref{fig:snap_timehist}(a), together with the capsule shape at the initial ($\dot{\gamma}_\mathrm{m} t$ = 0) and final states ($\dot{\gamma}_\mathrm{m} t$ = 50). The capsule, initially spherical, migrates towards the channel centerline while deforming, finally reaching its equilibrium position at the centerline, where it achieves an axial-symmetric shape. While the trajectory obtained with the highest frequency investigated ($f^\ast$ = 5) well collapses on that obtained with a steady flow, see Appendix~\S A
, when $f^\ast$ is small enough, the trajectory paths depend on the pulsation frequency, with the appearance of oscillations and with different axial migration speed.

The time history of the capsule deformation is shown in Fig.~\ref{fig:snap_timehist}(b), quantified by the Taylor parameter $D_{12} = |a_1 - a_2|/(a_1 + a_2)$, where $a_1$ and $a_2$ are the lengths of the semi-major and semi-minor axes of the capsule. Note that, we compute $D_{12}$ from the eigenvalues of the inertia tensor of an equivalent ellipsoid approximating the deformed capsule~\citep{Ramanujan1998}. The capsule deformation is maximized just after the flow onset when the capsule is subject to the high shear near the wall. As time passes, $D_{12}$ decreases and settles to a value which is around one order of magnitude smaller than the maximum (i.e., $O(D_{12}) = 10^{-2}$) when reaching the channel axis. 

\begin{figure}[htbp]
  \centering
    \includegraphics[clip,width=10cm]{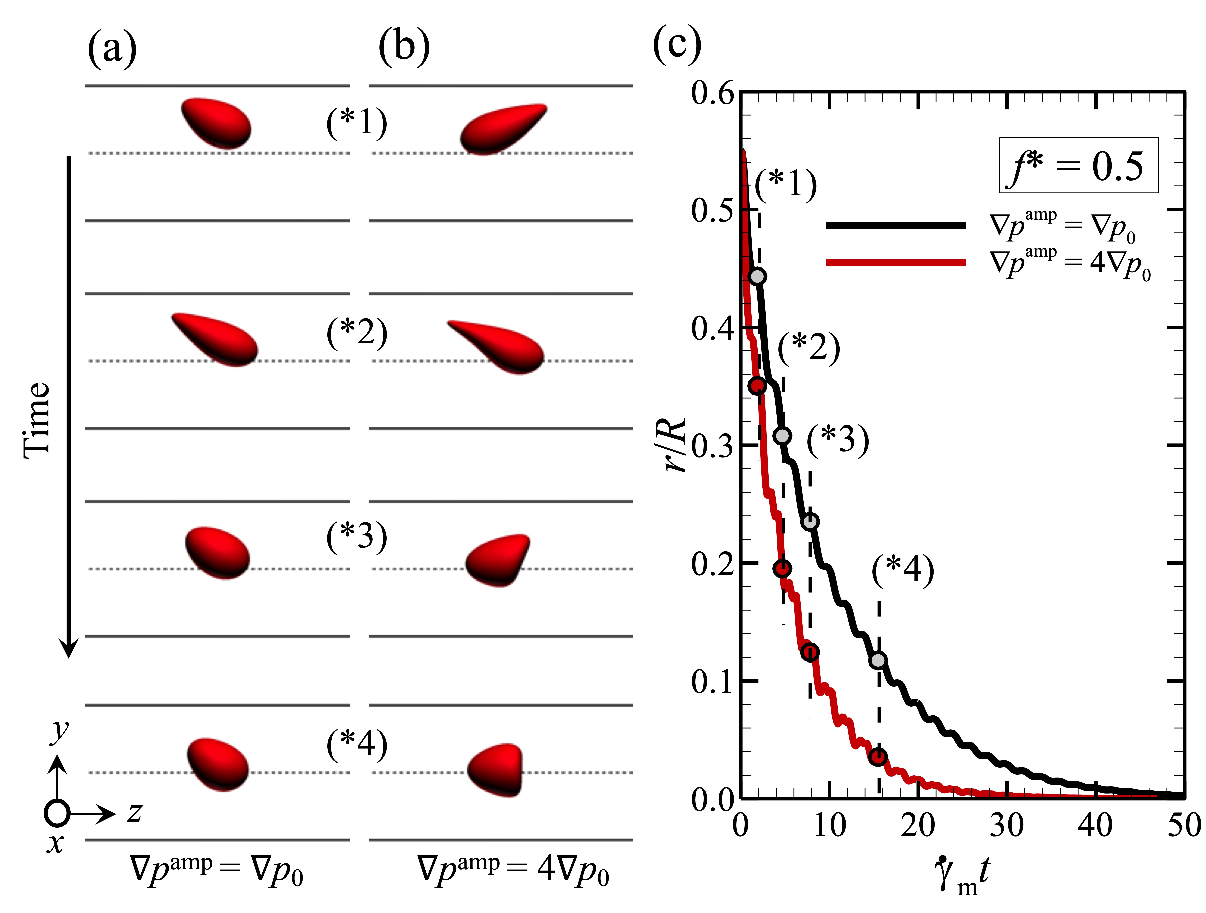}
    \caption{
    Side views of the capsule during its axial migration for $f^\ast = 0.5$ and different oscillatory amplitude: (a) $\nabla p^\mathrm{amp} = \nabla p_0$ and (b) $\nabla p^\mathrm{amp} = 4\nabla p_0$.
    The snapshots are taken at the time instants marked in (c), showed over the time history of $r/R$.
    All the results are obtained with $Ca = 1.2$, and $\lambda = 1$.
    }
    \label{fig:effect_amp}
\end{figure}

The migration speed is also affected by the amplitude of the oscillation $\nabla p^\mathrm{amp}$, as shown in Fig.~\ref{fig:effect_amp}, where the side views of the capsule during its axial migration for different $\nabla p^\mathrm{amp}$ (= $\nabla p_0$ and $4\nabla p_0$) are shown in Figs.~\ref{fig:effect_amp}(a) and \ref{fig:effect_amp}(b), respectively. The snapshots clearly show the capsule deformation and position as a consequence of the change of the background flow directions and oscillatory amplitudes. As $\nabla p^\mathrm{amp}$ increases, the capsule appears to migrate faster toward the channel centerline (Fig.~\ref{fig:effect_amp}c).

\begin{figure}[htbp]
  \centering
    \includegraphics[clip,width=17cm]{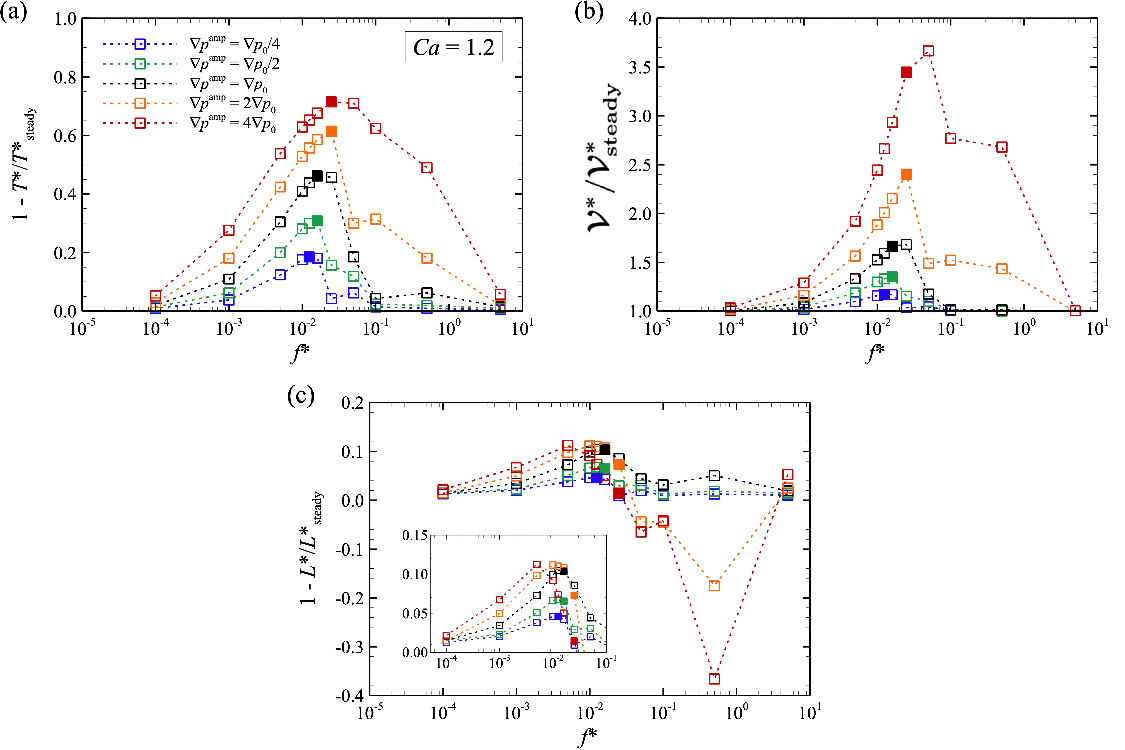}
    \caption{
    (a) The migration time $T^\ast$, (b) the migration speed $\m{V}^\ast$, and (c) the distance traveled during the migration $\m{L}^\ast$, normalized with those obtained in a steady flow ($T^\ast_\mathrm{steady}$, $\m{V}_\mathrm{steady}^\ast$, and $\m{L}^\ast_\mathrm{steady}$) as a function of $f^\ast$ and for different $\nabla p^\mathrm{amp}$.
    The results are obtained with $Ca$ = 1.2, and $\lambda$ = 1. The filled symbols in each panels represent the case with the optimal frequency which minimizes the migration time.
    }
    \label{fig:effect_f}
\end{figure}

To properly quantify the changes in axial migration, we define the migration time $T^\ast$ as the time needed by the capsule centroid to reach the centerline (within a distance of $\sim$6\% of its radius to account for the oscillations in the capsule trajectory). The ratio of the elapsed time $T^\ast$ and that in a steady flow is reported in Fig.~\ref{fig:effect_f}(a) as a function of $f^\ast$, for various pulsation amplitudes. The results clearly suggest that there exist a specific frequency to minimize the migration time. A very minor increase of the optimal frequency with the pulsation amplitude can be observed in the data. While the optimal frequency is almost independent of the pulsation amplitude, the migration time can be strongly reduced by its increase. Indeed, while the elapsed time is reduced by $18\%$ at the lowest amplitude investigated ($\nabla p^\mathrm{amp} = \nabla p_0/4$), it is reduced by $80\%$ at the highest one ($\nabla p^\mathrm{amp} = 4 \nabla p_0$).
Interestingly, the optimal frequency that minimizes the migration time ($O(f^\ast) = 10^{-2}$) is one order of magnitude smaller than the one which maximizes $D_{12}$ (Fig.~\ref{fig:d12} in Appendix~\S B), thus, suggesting that the axial migration time is unrelated to the maximum capsule deformation which happens in the initial stage of the capsule motion.

The changes in the migration time are clearly reflected in the migration speed $\m{V}^\ast = \m{V}/V_\mathrm{max}^\infty$, reported in Fig.~\ref{fig:effect_f}(b), which shows that when the migration time is minimum, the axial migration speed reaches almost its maximum. Here, the migration speed $\m{V}$ is defined as the ratio of the elapsed time $T$ and the traveled distance $\m{L}$ (i.e., $\m{V} = \m{L}/T$), defined as $\m{L} = \int_0^\m{L} |d\v{r}| = \int_0^\m{L} d\v{r} \cdot \hat{\v{t}} = \int_0^T \v{v} dt \cdot \hat{\v{t}}$, where $\hat{\v{t}} = \v{r}/|d\v{r}|$ is the unit tangential vector along the trajectory of the capsule centroid and $\v{v}$ is the the capsule centroid velocity.

The distance traveled by the capsule before completing the axial migration is reported in Fig.~\ref{fig:effect_f}(c) for the sake of completeness, showing that the optimal frequency to minimize the migration time, roughly corresponds to the minimization of the the traveled distance too.
Note that, the distance traveled during the migration $L^\ast$ depends not only on $f^\ast$ but also on $Ca$ (see Fig.~\ref{fig:effect_ca_on_L} Appendix~\S C).

\begin{figure}[htbp]
  \centering
    \includegraphics[clip,width=17cm]{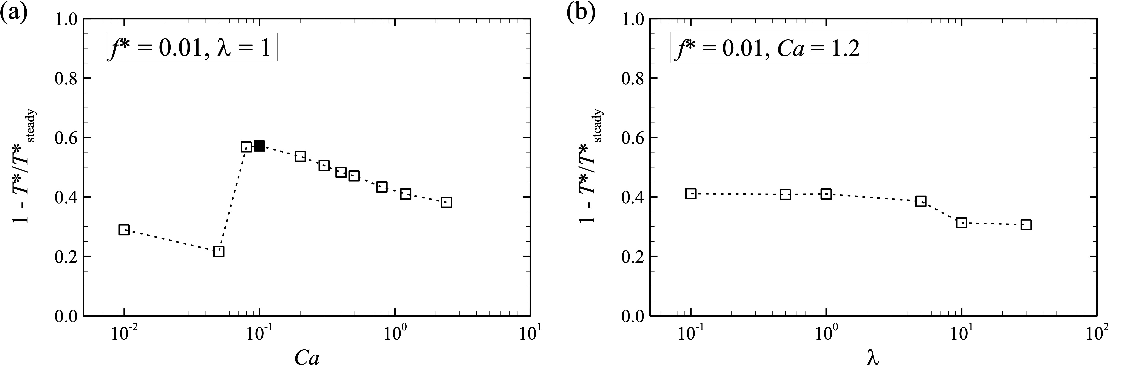}
    \caption{
    The migration time (a) as a function of $Ca$ at $\lambda = 1$ and $f^\ast = 0.01$ and (b) as a function of $\lambda$ at $Ca = 1.2$ and $f^\ast = 0.01$. The filled symbol in (a) represent the case with the optimal $Ca (= 0.1)$.
    }
    \label{fig:effect_ca_lambda}
\end{figure}

In summary, so far we have shown that, for a fixed $Ca$ and $\lambda$, there is an optimal frequency for the channel pulsation, able to minimize the capsule migration time by maximizing the migration speed and minimizing the traveled distance.
To complete our investigation, the effects of $Ca$ and $\lambda$ on the migration time $T^\ast$ are shown in Fig.~\ref{fig:effect_ca_lambda}. In particular, the results in Fig.~\ref{fig:effect_ca_lambda}(a) shows that the migration time depends on $Ca$, thus suggesting that the optimal frequency $f^\ast$ is also a function of $Ca$. On the other hand, as shown in Fig.~\ref{fig:effect_ca_lambda}(b), the migration time remains almost independent of the viscosity ratio for $\lambda \lesssim 5$.

\begin{figure}[htbp]
  \centering
    \includegraphics[clip,width=10cm]{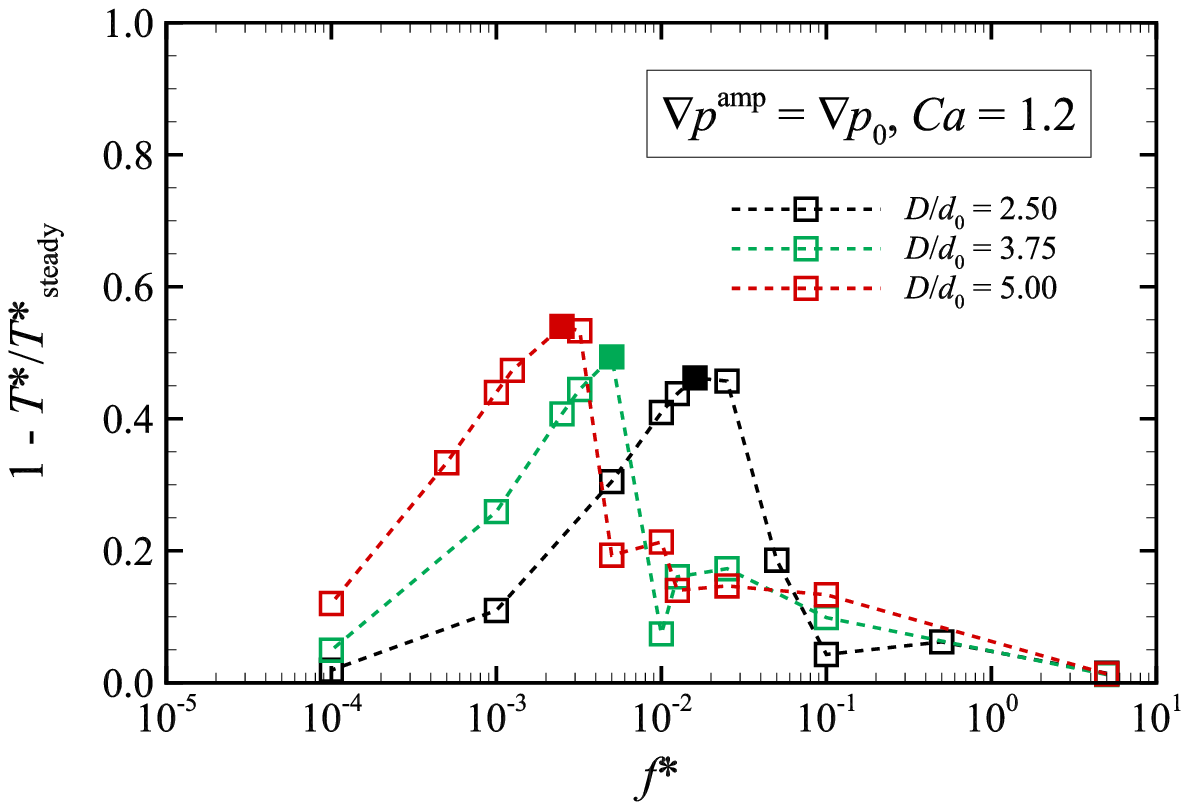}
    \caption{
    The ratio of the elapsed time $T^\ast$ to that in steady flow $T^\ast_\mathrm{steady}$ as a function of $f^\ast$.
    The these results are obtained with $Ca$ = 1.2, $\nabla p^\mathrm{amp}/\nabla p_0$ = 1, and $\lambda$ = 1.
    The filled symbols in each channel size ratio represent the case with the optical frequency which minimizes migration time.
    }
    \label{fig:effect_D}
\end{figure}

We also investigate the effect of the radial channel size on the migration time. Figure~\ref{fig:effect_D} shows the ratio of $T^\ast$ to that in a steady flow $T^\ast_\mathrm{steady}$ for two different channel size ratios $D/d_0$ = 2.5, 3.75, and 5, corresponding to $D (= 2R) = 20$ $\mu$m, $30$ $\mu$m, and $40$ $\mu$m for $d_0 (= 2a_0) = 8$ $\mu$m, as a function of the pulsation frequency $f^\ast$. For all cases, the initial position $r_0$ is set to be the same above (i.e., $r_0/R$ = 0.55). The results show that, independently of the channel size, the qualitative picture discussed above remains unchanged. While the amount of the speed-up of axial migration achieved with a pulsation remains almost unaltered (around $50\%$ for this case), the value of the optimal frequency changes with $D$, (the peak frequency reduces when $D$ is increased).

\section{IV. Conclusion}
In conclusion, we have proved that the axial migration speed of an elastic capsule in a pipe flow can be substantially accelerated by making the driving pressure gradient oscillating in time. We found that, the axial migration speed increases with the amplitude of the oscillation, while the most effective frequency revealed to be independent of the oscillatory amplitude. Also, we showed that the optimal frequency depends on $Ca$, but is basically independent of the viscosity ratio $\lambda$, overall proving that the changes in the axial migration are mostly due to the membrane elasticity.

The behaviour of capsules under pulsatile channel flows has been investigated in in some previous works~\citep{Lafzi2020, Maestre2019}. However, our study provides the first conclusive evidence of the acceleration of the axial migration of a capsule by pulsatile flow. Although it may be expected that the capsule is trapped in a state of resonance at the optimal frequency $f^\ast$ to minimize the migration time (Fig.~\ref{fig:effect_f}a), there is currently no clear theoretical framework on the resonance frequency of capsule in confined channel flows. Indeed, in our case the capsule configuration and its centroid are changing simultaneously, making the problem more complicated than what investigated in previous theoretical and numerical studies which assumed small deformations (i.e., weakly nonlinear problem) of drops~\citep{Chan1979, Magnaudet2003} and bubbles~\citep{Sugiyama2010}.

Given that the migration speed can be controlled by oscillatory frequency as well as background flow strength (or amplitude), the results obtained here can be utilised for label-free cell alignment/sorting/separation techniques, e.g., for circulating tumor cells in cancer patients or precious hematopoietic cells such as colony-forming cells. Our numerical results obtained physiologically relevant RBC properties in size $a_0$ and membrane elasticity $G_s$ form a fundamental basis for further studies on cellular flow mechanics in confined environments.


\section{Acknowledgments}
N.T. was supported by JSPS KAKENHI Grant Number JP20H04504. M.E.R. was supported by the Okinawa Institute of Science and Technology Graduate University (OIST) with subsidy funding from the Cabinet Office, Government of Japan. he presented study was partially funded by Daicel Corporation. N.T. thanks Dr. Naoto Yokoyama for helpful discussion. Finally, the collaborative research was supported by the SHINKA grant provided by OIST.

\appendix

\section{APPENDIX A: NUMERICAL VERIFICATION}\label{sec:appendix1}

\begin{figure}[htbp]
  \centering
    \includegraphics[clip,width=17cm]{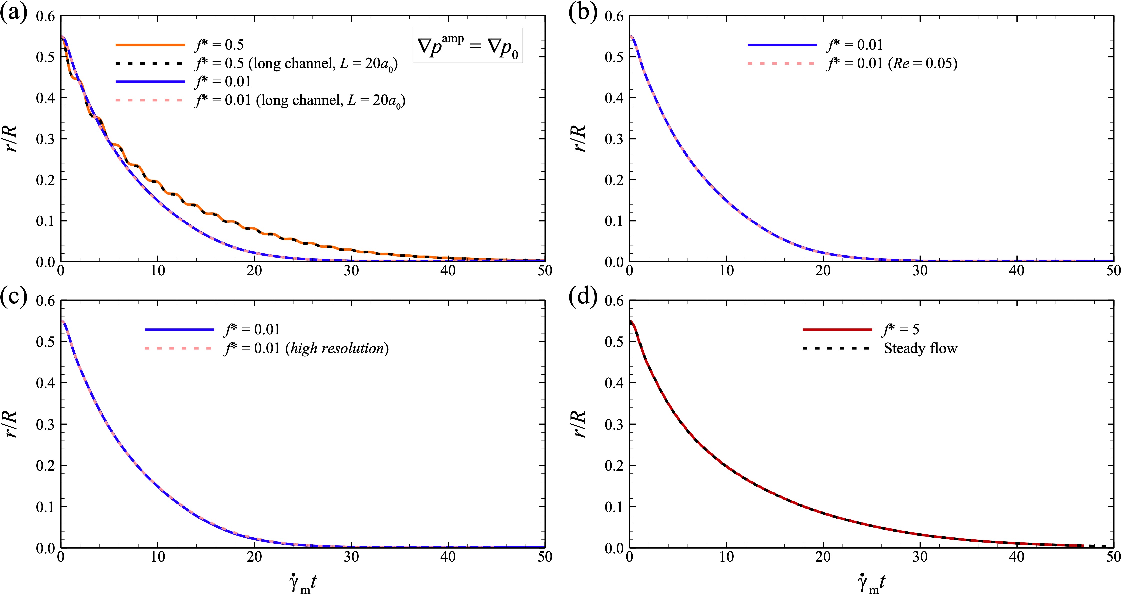}
    \caption{
    Time history of the radial position of the capsule centroid $r/R$ for (a) different channel lengths ($L = 10a_0$ and $20a_0$), (b) different Reynolds numbers $Re$ ($Re = 0.2$ and $0.05$), (c) different mesh resolutions ($250$ $\mu$m/lattice and $125$ $\mu$m/lattice).
    (d) Comparison of $r/R$ obtained with a steady flow and with the highest frequency investigated, $f^\ast = 5$.
    The results are obtained with $Ca = 1.2$, $\nabla p^\mathrm{amp} = \nabla p_0$, and $\lambda$ = 1.
    }
    \label{fig:verification}
\end{figure}

In this section, we provide additional verifications of the results provided in the main document. In particular, we investigate the effect of the channel length $L$, the Reynolds number $Re$, and the mesh resolutions on the trajectory of the capsule centroid, with the results reported in Fig.~\ref{fig:verification}(a)--\ref{fig:verification}(c). The figures show that no differences are observable when changing these parameters, thus suggesting that the domain is long enough, that the Reynolds number is small enough that our investigation can be considered in an inertialess condition, and that the numerical resolution is appropriate for the study. These results thus support the choice of parameters used for the rest of the investigation (i.e., $Re = 0.2$, $L = 10a_0$, and $250$ $\mu$m/lattice).

Finally, we show in Fig.~\ref{fig:verification}(d) that when the pulsation frequency is too large, the capsule does not experience the oscillatory flow. Indeed, the trajectory under the maximum frequency $f^\mathrm{\ast}$ investigated in this study well collapses on that obtained in a steady flow.

\section{APPENDIX B: THE MAXIMUM TAYLOR PARAMETER}\label{sec:appendix2}
\begin{figure}[htbp]
  \centering
    \includegraphics[clip,width=10cm]{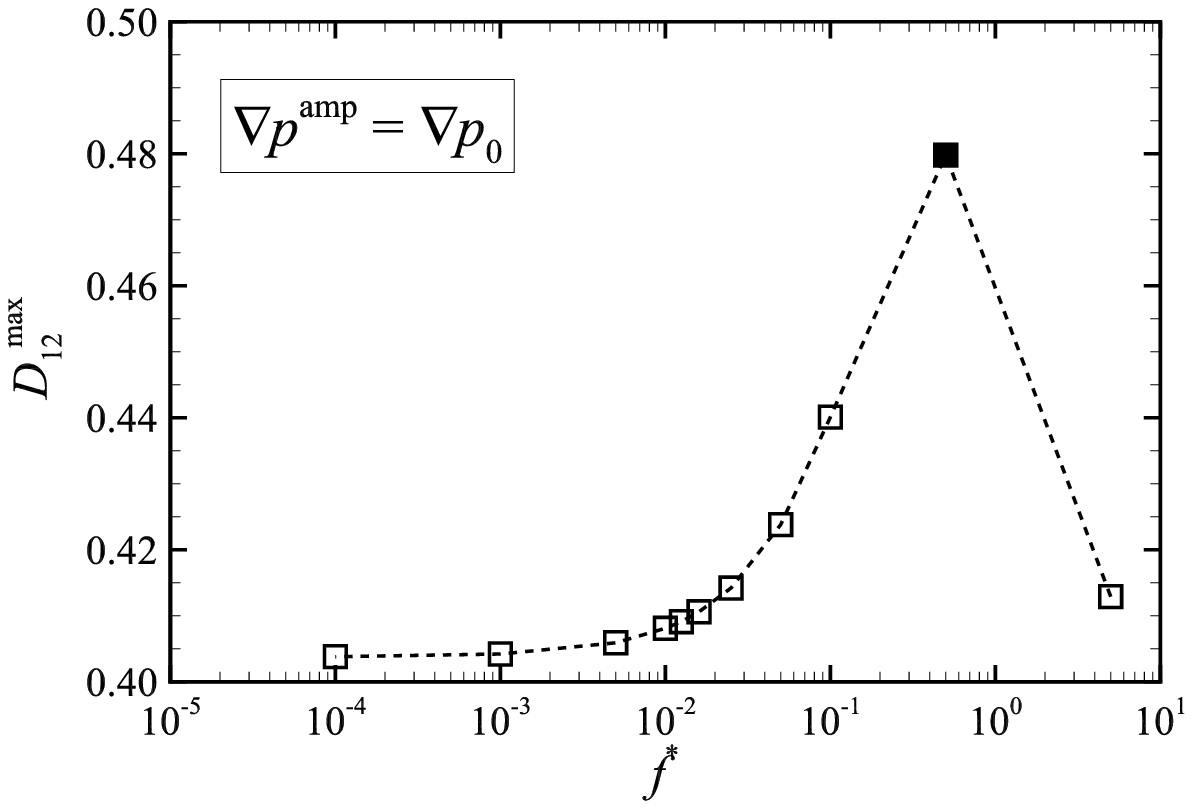}
    \caption{
    The maximum $D_{12}$ as a function of $f^\ast$.
    The filled symbol represents the case with the frequency which maximise $D_{12}$ after flow onsets.
    The results are obtained for $Ca = 1.2$, $\nabla p^\mathrm{amp} = \nabla p_0$, and $\lambda$ = 1.
    }
    \label{fig:d12}
\end{figure}

The maximum Taylor parameter $D_{12}^\mathrm{max}$, which can be observed just after the flow onset, is shown as a function of $f^\ast$ in Fig.~\ref{fig:d12}, for $Ca = 1.2$ and $\nabla p^\mathrm{amp} = \nabla p_0$. The result clearly shows that there is a specific $f^\ast$ which maximizes $D_{12}^\mathrm{max}$, which is higher than the optimal $f^\ast$ minimizing the migration time (Fig.~\ref{fig:effect_f}a).
\citet{Matsunaga2015} reported that at high frequency, a neo-Hookean spherical capsule undergoing oscillating sinusoidal shear flow cannot adapt to the flow changes and only slightly deforms according to predictions based on the asymptotic theory~\cite{BarthesBiesel1981, BarthesBiesel1985}. Thus, the capsule at low frequencies exhibits an overshoot phenomenon, in which the peak deformation is larger than its value in steady shear flow and increases with the viscosity contrast $\lambda$ and the mean value of $Ca$~\citep{Matsunaga2015}. Note that, our estimated frequency $f^\ast$ maximizing $D_{12}$ is one order magnitude smaller than that estimated by~\cite{Matsunaga2015}, difference that can be associated to the different membrane constitutive laws and flow profiles (i.e., simpler shear flow vs channel flow).

\section{APPENDIX C: EFFECT OF $Ca$ ON DISTANCE TRAVELED DURING THE MIGRATION}\label{sec:appendix3}
\begin{figure}[htbp]
  \centering
    \includegraphics[clip,width=10cm]{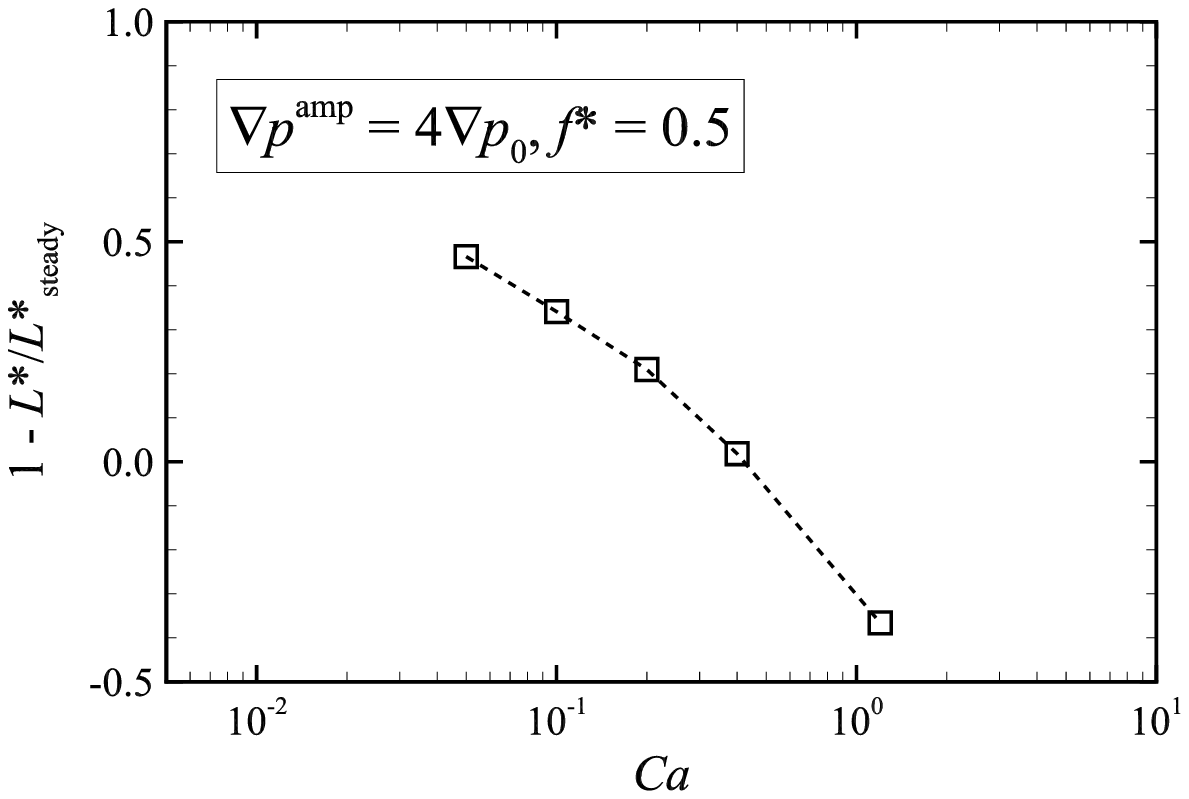}
    \caption{
    The distance traveled during the migration $L^\ast$, normalized with those obtained in a steady flow $L_\mathrm{steady}^\ast$ as a function of $Ca$ for $\nabla p^\mathrm{amp} = 4 \nabla p_0$ and $f^\ast = 0.5$. The results are obtained with $\lambda = 1$.
    }
    \label{fig:effect_ca_on_L}
\end{figure}

From Fig.~\ref{fig:effect_f}(c), it seems that the amplitude of oscillation can decrease significantly the relaxation process in some cases. To confirm whether this effect is robust with respect to $Ca$, we investigate the distance traveled during the migration $L^\ast$ for different $Ca$ ($= 0.05, 0.1, 0.2$ and $0.4$) with $\nabla p^\mathrm{amp} = 4\nabla p_0$ and $f^\ast = 0.5$, when $L^\ast$ tends to be longer that in the steady flow (i.e., $1 - L^\ast/L_\mathrm{steady}^\ast < 0$). 
From the results in Fig.~\ref{fig:effect_ca_on_L}, we can observe that the travel distance $L^\ast$ is longer than in steady flow only for high $Ca (\geq 1.2)$.

\bibliography{references}

\end{document}